\documentclass[prc,showpacs
,floatfix,letterpaper,superscriptaddress%
,preprint%
]{revtex4}
\usepackage{graphicx,subfigure}
\usepackage{comment}
\usepackage{amsmath, amsthm, amssymb}
\usepackage{bm}
\usepackage{color}

\def\nuc#1#2{\relax\ifmmode{}^{#1}{\protect\text{#2}}\else${}^{#1}$#2\fi}

\begin{document}

\title{Optical model potential of deuteron with $1p$-shell nuclei}

\author{Y. Zhang}
\affiliation{School of Physics and State key laboratory of nuclear physics and technology, Peking University, 100871, Beijing}

\author{D.Y. Pang}
\email[Correspondence author: ]{dypang@buaa.edu.cn}
\affiliation{School of Physics and Nuclear Energy Engineering, Beihang University, Beijing 100191, China}
\affiliation{Beijing Key Laboratory of Advanced Nuclear Materials and Physics, Beihang University, Beijing 100191, China}

\author{J.L. Lou}
\email[Correspondence author: ]{jllou@pku.edu.cn}
\affiliation{School of Physics and State key laboratory of nuclear physics and technology, Peking University, 100871, Beijing}

\date{\today}


\begin{abstract}
A set of global optical potential parameters, DA1p, for deuterons with the $1p$-shell nuclei is
obtained by simultaneously fitting 67 sets of experimental data of deuteron elastic scattering from \nuc{6}{Li},
\nuc{9}{Be}, \nuc{10}{B}, \nuc{11}{B}, \nuc{12}{C}, \nuc{13}{C}, \nuc{14}{N}, \nuc{16}{O}
 and \nuc{18}{O} with incident energies between
5.25 and 170 MeV.
DA1p improves the description of the deuteron elastic scattering
from the $1p$-shell nuclei with respect to the existing systematic deuteron potentials and can give satisfactory reproduction
to the experimental data with radiative nuclei such as \nuc{9}{Li}, \nuc{10}{Be}, \nuc{14}{C} and \nuc{14}{O}.
\end{abstract}

\maketitle


\section{Introduction}

Systematic optical model potentials (OMPs) are very useful in many fields of nuclear physics. They help to
reduce the uncertainties of nuclear structure information extracted from direct nuclear reactions \cite{LiuXD-PRC-2004}
and to make systematic analyses \cite{Tsang-PRL-2005, Tsang-PRL-2009}. They are also indispensable in reliable predictions of
reaction rates of direct nuclear reactions which are not easy or impossible to be measured directly in laboratories.
Over the past several decades, many systematic potentials have been proposed
for nucleon ($A=1$) \cite{BGPN, CH89, KD02, XuRR-PRC-2012}, deuteron ($A=2$) \cite{Perey-PRev-1963, Lohr-NPA-1974, Daehnick, HanYL-PRC-2006,AnHX-PRC-2006},
\nuc{3}{H} and \nuc{3}{He} ($A=3$) \cite{BG3He, GuoHR-PRC-2009, GDP08, LiXiaohua, LiangChuntian}, alpha-particles
($A=4$) \cite{Nolte-PRC-1987,Avrigeanu-ADDNT-2009, Pang4He, Pang-JPG-2012}, and heavy ions ($A\geq6$) \cite{Chamon-PRC-2002, Furumoto-PRC-2012, XuYP-PRC-2013}.
They are widely used in studies of direct nuclear reactions.

In three-body models of the $(d,p)$ and $(p,d)$ reactions proton and neutron
potentials are used instead of the deuteron-target potentials \cite{Johnson-Soper,Akram-PRC-2012,Pang-PRC-2013}. The deuteron-target OMPs are necessary in
many other reactions, for instance, the $(\nuc{3}{He},d)$ reactions, for which the distorted wave Born approximation is usually still valid \cite{Vernotte-NPA-1994,Jenny-CPL-2014}.
Deuteron optical potentials are also needed in reactions induced by radiative nuclei with a deuterium target in inverse kinematics.
In such reactions, one usually focuses on the weakly-bound nature of the radiative nuclei and will need the deuteron potential with the core nucleus
in calculations with the continuum discretized coupled channel (CDCC) method \cite{ChenJie-PRC-2016}.

Most of the existing systematic deuteron potentials are based on the analysis of
angular distributions of elastic scattering cross sections of deuterons from
heavy targets with atomic masses of, typically, $A_\textrm{T}\gtrsim30$ \cite{Perey-PRev-1963, Lohr-NPA-1974, Daehnick}.
It is well-known that the systematics developed at such heavy mass region are
different from that in the light mass region ($A_\textrm{T}\lesssim20$) \cite{Watson-PR-1967, HT1p}.
Phenomenological renormalization factors are found to be needed when systematic potentials developed in the heavy-mass region
are applied to light targets \cite{Faisal-CPL-2010,ChenJie-PRC-2016}.  This is not convenient in theoretical calculations
for reactions which have no corresponding elastic scattering data to constrain the OMP parameters. The database for the systematics of
An Haixia \textit{et al.} and Han YinLu \textit{et al.} include \nuc{12}{C}, \nuc{14}{N}, and \nuc{16}{O} targets \cite{HanYL-PRC-2006,AnHX-PRC-2006}.
However, their systematics are not optimized for the light-target region. Many experimental data with other light-heavy targets are not included. It is thus very useful
to establish a systematic deuteron potential for the $1p$-shell nuclei.

In this paper, we report a systematic deuteron potential with the $1p$-shell nuclei. It is designated as DA1p. The target nuclei include \nuc{6}{Li},
\nuc{9}{Be}, \nuc{10}{B}, \nuc{11}{B}, \nuc{12}{C}, \nuc{13}{C}, \nuc{14}{N}, \nuc{16}{O},
 and \nuc{18}{O} with deuteron incident energies between
5.25 and 170 MeV. The experimental data available for the \nuc{6}{Li} and \nuc{7}{Li} targets are mostly limited
for deuteron energies below 14.7 MeV. Within such a low energy region, contributions from the compound processes are expected
to be important in the elastic scattering of deuterons with these lightest $1p$-shell nuclei. For this reason, these data
are analyzed individually. The parameterization of DA1p is described in Section \ref{sec-parameterization}; the resulting
OMP parameters are reported in Section \ref{sec-results} with comparisons
between optical model calculations and experimental data. Examination of the application of DA1p to the total cross sections and elastic scattering from radiative nuclei are shown in Section \ref{sec-discussions}.  Our conclusions are given in Section \ref{conclusions}.

\section{Parameterization and determination of the systematic potential parameters}

\subsection{Parameterization}\label{sec-parameterization}

The parameterization of the optical model potential  in this work, $U(r)$, as a function of $r$ which is the distance between a projectile and
target nuclei, is the same as that of HT1p \cite{HT1p}:
\begin{eqnarray}\label{eq-form-of-op}
 U(r) &=& -V_{\textrm{v}}f_{\textrm{ws}}(r,R_\textrm{r},a_\textrm{r})-iW_{\textrm{v}}f_{\textrm{ws}}(r,R_\textrm{w}, a_\textrm{w})\nonumber\\
& & -iW_{\textrm{s}}(-4a_\textrm{w})\frac{d}{dr}f_{\textrm{ws}}(r,R_\textrm{w}, a_\textrm{w})\nonumber\\
& & + V_{\textrm{C}}(r),
\end{eqnarray}
where $V_{\textrm{v}}$, $W_{\textrm{v}}$, and $W_{\textrm{s}}$ are the depths of the real, and the volume and surface imaginary potentials, respectively.
$f_{\textrm{ws}}(r)$ is the Woods-Saxon form factor:
\begin{equation}
 f_{\textrm{ws}}(r,R_i,a_i)=\frac{1}{1+\exp{[(r-R_i)/a_i]}},
\end{equation}
with $i=$v and w for the real and imaginary potentials, respectively.

The diffuseness of these potentials are assumed to be independent on the target masses ($A_\textrm{T}$) and incident energies ($E$ in MeV).
Such dependences, however, are parameterized in the radius parameters:
\begin{equation}
R_i=r_iA_{\textrm{T}}^{1/3}+r_{i}^{(0)} + r_{i\textrm{e}}(E-E_{\textrm{C}}),
\end{equation}
where $E_{\textrm{C}}$ is the Coulomb correction to the incident energy \cite{Perey-PRev-1963, CH89,GDP08}:
\begin{equation}\label{eq-Ec}
 E_{\textrm{C}}=\frac{6Z_{\textrm{P}}Z_{\textrm{T}}e^2}{5R_{\textrm{C}}},
\end{equation}
in which $Z_{\textrm{T}}$ and $Z_{\textrm{P}}$ are the charge numbers of the target and the projectile nuclei, respectively, and
$R_\textrm{C}=r_\textrm{c}A_\textrm{T}^{1/3}$ is the radius of the Coulomb potential:
\begin{equation}
V_{\textrm{C}}(r)=
\begin{cases}
 \frac{\displaystyle Z_{\textrm{P}}Z_{\textrm{T}}e^2}{\displaystyle r},
    & (r\geqslant R_{\textrm{C}})\\
 \frac{\displaystyle Z_{\textrm{P}}Z_{\textrm{T}}e^2}{\displaystyle
2R_{\textrm{C}}}\left(3-\frac{\displaystyle
r^2}{\displaystyle R_{\textrm{C}}^2}\right)  & (r\leqslant R_{\textrm{C}}).
\end{cases}
\end{equation}
In this work $r_{\textrm{c}}$ is fixed to be 1.3 fm. The energy dependence of the radius parameters
for both the real and the imaginary parts are found to be important
to simultaneously describe the elastic scattering data in a wide energy range, as suggested in Ref.\cite{PangandYe-PRC-2011}.

The depth of the real potential in Eq.(\ref{eq-form-of-op}) is assumed to depend linearly on the incident energies:
\begin{equation}\label{eq-Vr}
 V_{\textrm{v}}(E) = V_{\textrm{r}} + V_{\textrm{e}}(E-E_{\textrm{C}}),
\end{equation}
The volume and surface terms of the imaginary potentials, $W_\textrm{v}$ and $W_\textrm{s}$, are defined as
\begin{eqnarray}\label{eq-parameterization-imag}
 W_{\textrm{v}}(E) &=& \frac{W_{\textrm{v0}}}{1+\exp\left(\frac{W_{\textrm{ve0}}-(E-E_{\textrm{C}})}{W_{\textrm{vew}}}\right)},\\
 W_{\textrm{s}}(E) &=& \frac{W_{\textrm{s0}}}{1+\exp\left(\frac{(E-E_{\textrm{C}})-W_{\textrm{se0}}}{W_{\textrm{sew}}}\right)}.
\end{eqnarray}
For \nuc{6,7}{Li} at low energies, the imaginary potentials are assumed to depend linearly on the incident energies: $W_j=W_{j0}+W_{je}(E-E_C)$,
with $j=$v and s for the volume and surface imaginary parts, respectively.

Spin-orbit potentials are not included in the parameterization of DA1p. We do this because of two practical reasons. Firstly, we expect DA1p to be used in
CDCC calculations for reactions induced by weakly-bound radiative nuclei with the deuterium targets. Currently, in most CDCC calculations using, for example, computer code FRESCO \cite{FRESCO},
spin-orbit couplings are not implemented. In such cases we need the OMP of deuteron to reproduce the elastic scattering data without a spin-orbit potential as well. Secondly, the experimental data analyzed in this work are all angular distributions of elastic scattering cross sections, which are not sensitive to the spin-orbit potentials,
especially at forward angles, where the data are most well accounted for by the optical model. In total we have 16 free parameters for DA1p,
which are listed in Table.\ref{tab-1p-parameters}.

\subsection{Parameters of DA1p and comparisons with experimental data}\label{sec-results}

89 sets of experimental data for deuteron elastic scattering from the $1p$-shell nuclei are analyzed in this work,
which consist of 65 sets for \nuc{9}{Be}, \nuc{10,11}{B}, \nuc{12,13}{C},
\nuc{14}{N}, and \nuc{16,18}{O} with incident energies below 171 MeV and 24 sets for \nuc{6,7}{Li} from 4.5 MeV to 171 MeV.
All the data sets are obtained from the EXFOR database \cite{exfor}. Details of these data are shown in Table.\ref{tab-1p-data}.

\begin{table*}[htbp]
\caption{The database used in searching of the parameters of DA1p. $\chi^2$ values with the systematics of DA1p, Daehnick $el\ at$, and Haixia An $et\ al$, are labeled as  $\chi^2_{\textrm{DA1p}}$ $\chi^2_{\textrm{Dae}}$ and $ \chi^2_{\textrm{An}}$, respectively. The unit of $E_\textrm{d}$ is MeV.}
\begin{scriptsize}
\begin{center}
\begin{tabular}{cccccccccccccccccccccc}
\hline\hline
target & $E_\textrm{d}$ & $\chi^2_{\textrm{DA1p}}$ & $\chi^2_{\textrm{Dae}}$ &$ \chi^2_{\textrm{An}}$ & Ref & target & $E_\textrm{d}$ & $\chi^2_{\textrm{DA1p}}$ & $\chi^2_{\textrm{Dae}}$ & $\chi^2_{\textrm{An}}$& Ref &target & $E_\textrm{d}$ & $\chi^2_{\textrm{DA1p}}$ & $\chi^2_{\textrm{Dae}}$ & $\chi^2_{\textrm{An}}$& Ref \\ \hline

\nuc{6}{Li} &  4.5 &  0.46 & 39.09 & 16.47 &  \cite{POWELL-NPA-1970} & \nuc{9}{Be} & 7 & 1.69 & 20.30 & 19.93 &  \cite{SZCZUREK-ZPA-1989} & \nuc{12}{C} &25.9 & 6.48& 8.81& 8.79 &  \cite{Dantzig-NP-1963}\\

& 4.75 & 1.55 & 40.02 & 16.60 &\cite{POWELL-NPA-1970} & & 7.5 & 2.16 & 11.99 &10.94 &  \cite{Generalov-IZV-2000} &  & 29.5 & 9.87 & 5.55& 5.03 &\cite{Perrin-NPA-1977}\\

& 5 &  1.72 & 56.91& 24.45  & \cite{POWELL-NPA-1970}&  & 8 & 2.35 & 13.44 & 11.58 &\cite{Generalov-IZV-2000} & & 34.4 & 1.30 & 1.39 &1.94 & \cite{Newman-NPA-1967} \\

& 5.25 & 1.83 & 49.40 &21.29 & \cite{POWELL-NPA-1970} &  & 8.5 & 3.98 & 19.10 & 16.02 & \cite{Generalov-IZV-2000}  & & 52  & 0.88 & 1.31 & 5.34 &\cite{Hinterberger-NPA-1968} \\

& 6 & 6.16 & 57.75 & 25.71& \cite{Abramovich-IZV-1976}  & & 9 & 2.45 & 11.30 & 9.24 & \cite{Generalov-IZV-2000}  & &  53.3 & 0.80 & 2.59 & 3.25 & \cite{Ishida-PLB-1993}\\

& 7 & 5.05 & 58.69 & 26.08 &  \cite{Abramovich-IZV-1976}  &  & 9.5 & 2.97 & 13.89 & 11.64 &\cite{Generalov-IZV-2000} & & 56 & 2.45 & 3.50 & 10.89 &\cite{Matsuoka-NPA-1986} \\

& 8 & 3.49 & 64.53  & 28.46 & \cite{Abramovich-IZV-1976} &  & 11.8 & 7.50 & 42.43 & 40.14 & \cite{Fitz-NPA-1967} &  & 60.6 & 3.04 & 10.12 & 4.36 & \cite{Aspelund-NPA-1975} \\

& 8 & 3.02 & 58.13 & 25.64 &  \cite{BINGHAM-NPA-1971} & & 13.6 & 8.52 & 23.67 & 25.09 &\cite{Vereshchagin-SPJ-1970} &  & 77.3 & 0.72 & 8.48& 5.31 & \cite{Aspelund-NPA-1975}\\

& 9 & 1.21 & 63.41& 27.65 &  \cite{Abramovich-IZV-1976}  &  & 15 & 10.62  &42.04 & 40.70 &  \cite{DARGEN-NPA-1976}&& 80 & 0.68 & 3.64 & 1.45 & \cite{Duhamel-NPA-1971} \\

& 10 & 0.45 & 60.86  & 26.34 &  \cite{Abramovich-IZV-1976}  & & 15.8 & 8.04 & 30.24 &  30.24 &\cite{COWLEY-NP-1966}   &   & 90 & 0.60 & 29.07 & 7.83 &\cite{Aspelund-NPA-1975}\\

& 10 & 0.90 & 65.81 & 28.73&   \cite{BINGHAM-NPA-1971} & & 24 & 4.99 & 22.19 & 23.53 & \cite{SUMMERSGILL-PRev-1958}  && 110 & 0.58 & 12.36 & 6.78 &\cite{Betker-PRC-1993} \\

&11.8 & 1.63 & 63.88 & 27.65 &  \cite{Ludecke-NPA-1968}   & & 27.7 & 4.19 & 3.77 & 5.44 &  \cite{Slobodrian-NP-1962} &  & 120 & 1.13 & 6.39 & 10.04 &\cite{Betker-PRC-1993}\\

& 12 & 2.07 & 66.42 & 28.98 & \cite{BINGHAM-NPA-1971} & \nuc{10}{B} & 11.8 & 4.55 & 11.93 & 12.03 & \cite{Fitz-NPA-1967}   &  & 140 & 3.86 & 28.83 & 15.97 &  \cite{Okamura-PRC-1998}\\

& 14.7 & 5.40 & 44.40 & 20.54 &\cite{MATSUKI-JPJ-1969} & \nuc{11}{B} & 11.8 & 3.47 & 41.45 & 35.09 &\cite{Fitz-NPA-1967}  &  &170 & 4.75 & 10.22 & 13.61 &\cite{Baeumer-PRC-2001} \\

& 25 & 3.73 & 5.62 &  6.04 &\cite{Burtebyaev-PAN-2010}  & & 13.6 & 2.87 & 21.64 & 19.52 &  \cite{Vereshchagin-SPJ-1970} &  \nuc{13}{C} &13.7 &  7.89 & 15.85 & 19.38 &\cite{Guratzsch-NPA-1970} \\

&171 & 1.41 & 13.91 & 2.67 & \cite{Korff-PRC-2001} &  & 27.7 & 4.41 & 2.93 & 2.81 & \cite{Slobodrian-NP-1962} &   & 17.7 & 2.92 & 18.69 & 13.41 & \cite{Peterson-NPA-1984} \\

\nuc{7}{Li} & 6 & 2.96 & 66.64 & 29.79 & \cite{Abramovich-IZV-1976}   & \nuc{12}{C} & 7 & 2.43 & 13.91 & 13.98 & \cite{Ohlsen-NP-1963} & \nuc{14}{N} & 11.8 & 2.15 & 10.40 & 8.34 & \cite{Fitz-NPA-1967} \\

& 7 & 1.86 & 68.97 & 30.67 & \cite{Abramovich-IZV-1976}  & & 8 & 1.44 & 17.15 & 14.32  & \cite{Ohlsen-NP-1963} & & 52 & 2.65 & 2.91 & 7.62 & \cite{Hinterberger-NPA-1968} \\

& 8 & 0.93 & 69.63 & 30.63 &  \cite{Abramovich-IZV-1976} & & 9& 2.81 & 12.49 &  11.31 & \cite{Ohlsen-NP-1963} & \nuc{16}{O}& 5.25 & 1.25 & 5.71 & 6.51 & \cite{Davison-CJP-1970}  \\

& 9 & 0.64 & 70.55 &  30.93 & \cite{Abramovich-IZV-1976} & & 10.6 & 3.23 & 13.33 &  10.67 & \cite{BALDEWEG-NP-1966} &  & 5.5 & 0.63 & 4.57 & 5.21& \cite{Davison-CJP-1970}  \\

& 10 & 0.81 & 74.14 &  32.32 & \cite{Abramovich-IZV-1976} & & 11 & 5.37 & 13.81 & 13.56 &\cite{BALDEWEG-NP-1966} & & 6 & 2.83 & 5.46 & 6.06& \cite{Davison-CJP-1970}\\

&11.8 & 1.21 & 77.55  & 33.48 & \cite{Ludecke-NPA-1968} & & 11& 3.84 & 18.79 & 16.28 & \cite{Ohlsen-NP-1963} &  & 11.8 & 1.51 & 13.00 & 8.37 &\cite{Fitz-NPA-1967} \\

& 12 & 1.23 & 62.03 & 26.50 & \cite{BINGHAM-NPA-1971} & & 11.4 & 3.77 & 15.08 & 12.36  & \cite{BALDEWEG-NP-1966} &   & 13.3 & 6.61 & 7.20 & 7.08 & \cite{Corrigan-NPA-1972}\\

& 14.7 & 1.17 & 53.22 & 23.41 &\cite{MATSUKI-JPJ-1969} & & 11.8 & 4.53 & 33.65 & 26.53 & \cite{Fitz-NPA-1967} &  & 15.8 & 6.78 & 9.63 & 9.53 & \cite{COWLEY-NP-1966}\\

\nuc{9}{Be} & 5.5 & 1.70 & 11.98 & 10.73 & \cite{POWELL-NPA-1970}  &  & 12.4 & 3.28 & 10.77 & 10.13 &\cite{BALDEWEG-NP-1966} & & 34.4 & 3.98 & 4.55 & 5.72 &  \cite{Newman-NPA-1967}\\

& 5.75 & 0.95 & 11.50 & 9.90 & \cite{POWELL-NPA-1970}  & & 12.8 & 6.57 & 14.00 & 14.94 &   \cite{BALDEWEG-NP-1966} &  & 52 & 1.92 & 8.49 & 2.65 & \cite{Hinterberger-NPA-1968}\\

&6 & 2.97 & 7.09 & 4.76 &\cite{Generalov-IZV-2000} & & 13.2 & 6.63 & 19.94 & 21.80 &    \cite{BALDEWEG-NP-1966} &  & 56 & 3.64 & 1.63 &  4.61 &\cite{Hatanaka-NPA-1980}\\

& 6 & 4.34 & 4.70 & 8.41 & \cite{POWELL-NPA-1970}  & & 13.9 & 6.62 & 15.18 & 17.96 &  \cite{BALDEWEG-NP-1966} &  & 171 & 4.78 & 3.06 & 14.39 & \cite{Korff-PRC-2001} \\

& 6.5 & 2.67 & 7.95 & 8.54 &\cite{Generalov-IZV-2000} & & 15.3 & 11.20 & 59.22 & 54.36 & \cite{Galanina-PAN-2007}   & \nuc{18}{O} & 16.3 & 1.62 & 7.25 & 7.64 &\cite{Burjan-JP}\\

& 7 & 1.69 &  12.29 & 11.76 &\cite{Generalov-IZV-2000} & & 15.8 & 7.63 & 37.28 & 32.93 & \cite{COWLEY-NP-1966} \\
\hline
\end{tabular}
\end{center}
\end{scriptsize}
\label{tab-1p-data}
\end{table*}

In searching for the parameters of DA1p, 65 sets of data for the \nuc{9}{Be}, \nuc{10,11}{B}, \nuc{12,13}{C},
\nuc{14}{N}, and \nuc{16,18}{O} targets together with the two sets of \nuc{6}{Li} at 25 and 171 MeV are simultaneously fitted using the computer code MINOPT \cite{CH89}.
The OMP parameters are optimized with the usual minimization of $\chi^2$ method:
\begin{equation}\label{eq-Qsquare-def}
  \chi^2 =\frac{1}{N}\sum_{i=1}^N\frac{\left[\sigma_{i}^{\text{exp}}-\sigma_{i}^{\text{th}}\right]^2}{\Delta\sigma_{i}^2},
\end{equation}
where $\sigma_{i}^{\text{exp}}$, $\sigma_{i}^{\text{th}}$, $\Delta\sigma_{i}$ are the experimental and theoretical cross sections, and the experimental errors, respectively. Uniform uncertainty of the experimental data $\Delta\sigma_{i}/\sigma_{i}^\text{exp}=15\%$ is assumed in this work.
Experimental data measured by different groups may have different systematic uncertainties. For this reason, normalization of the experimental data
is allowed during the parameter searching with MINOPT \cite{CH89}.
Convergence of the searching is ensured by observing that the values of parameters came back to their optimized ones (within their uncertainties) in fittings with different initial values randomly set within 10\%.
The uncertainties of the parameters of DA1p are obtained with the bootstrap method \cite{Efron-bootstrap},
which reduplicates the calculations 1000 times by random sampling with replacements of the datasets used in the original database.
Details of applying the bootstrap method for the uncertainties of the systematic OMPs can be found in Refs.\cite{CH89, GDP08}. The final parameters of DA1p and their uncertainties are given in Table.\ref{tab-1p-parameters}.

During this work, we found that the experimental data of \nuc{6}{Li} and \nuc{7}{Li} at low incident energies ($E<15$ MeV) are hard to be described together with the other datasets using a systematic potential. These data may have considerable contributions from compound processes, which can not be accounted for by the optical model implemented in MINOPT.
We search for the deuteron potentials with \nuc{6}{Li} and \nuc{7}{Li} at low energies separately.
The results are also given in Table.\ref{tab-1p-parameters}. As one can see, the parameters of these two targets differ very much from the systematics established by the other $1p$-shell nuclei.

\begin{table}[htbp]
\scriptsize
\caption{Values of parameters, $P$, and their uncertainties, $\Delta P$, of DA1p.
$V_{\textrm{r}}$, $V_{e}$, $W_\textrm{v0}$, $W_\textrm{s0}$, $W_\textrm{se}$, $W_\textrm{ve0}$, $W_\textrm{vew}$, $W_\textrm{se0}$, and $W_\textrm{sew}$ are in MeV, and  $r_{\textrm{r}}$, $r_{\textrm{r}}^{(0)}$, $r_\textrm{re}$, $a_{\textrm{r}}$, $r_\textrm{w}$, $r_\textrm{w}^{(0)}$,$r_\textrm{we}$, and $a_\textrm{w}$ are in femtometers.}
\begin{center}
\begin{tabular}{ccccccc}
\hline\hline
 & \multicolumn{2}{c}{$1p$-shell} & \multicolumn{2}{c}{$^{6}$Li} & \multicolumn{2}{c}{$^{7}$Li} \\ \hline
parameter & $P$    &$\Delta P$ & $P$   &  $\Delta P$ & $P$    & $\Delta P$ \\  \hline
$V_{\textrm{r}}$             & 98.9  & 0.8                  &47.9   & 1.5                    & 26.1  & 0.7  \\
$V_{\textrm{e}}$                           & -0.278 & 0.021             & 2.37 & 0.17                   & 1.19  & 0.02 \\
$r_{\textrm{r}}$             & 1.11  & 0.001                & 1.62  & 0.01                 & 1.45    & 0.01 \\
$r_{\textrm{r}}^{(0)}$     & -0.172  & 0.004            & -     & -                           &  -  & - \\
$r_\textrm{re}$             & 0.00117 &0.00009        & -0.0122     & 0.0021     & 0.097     & 0.005    \\
$a_{\textrm{r}}$            & 0.776  & 0.001             & 0.876 & 0.037               & 0.844   & 0.008 \\
$W_\textrm{v0}$          & 11.5   & 0.7                   & -     & -                           & -   & -  \\
$W_\textrm{s0}$          & 7.56   & 0.67                 & 11.3    & 0.8                  & 215.0   & 0.2  \\
$r_\textrm{w}$             & 0.561   & 0.006             & 2.83  & 0.039               & 2.12  & 0.02 \\
$r_\textrm{w}^{(0)}$    &3.07   &0.002                  & -     & -                           &  - & - \\
$r_\textrm{we}$           & -0.00449 & 0.00008     & -0.0911     & 0.0002         & 0.022      & 0.003  \\
$a_\textrm{w}$            & 0.744  & 0.001               & 0.27 & 0.004               & 0.261   & 0.104 \\
$W_\textrm{se}$         & -   & -                             & 3.44   & 0.32                & -16.1   & 0.1  \\
$W_\textrm{ve0}$       & 18.1  & 1.4                      & -     & -          & -  & - \\
$W_\textrm{vew}$      & 5.97   & 1.33                    & -     & -          & -   & -  \\
$W_\textrm{se0}$      &  14.3   & 1.4                      & -  & -          & -   & - \\
$W_\textrm{sew}$      & 4.55  & 0.86                   & - & -          & -  & -  \\ \hline
\end{tabular}
\end{center}
\label{tab-1p-parameters}
\end{table}

\begin{figure}[htbp]
 \includegraphics[width=0.5\textwidth]{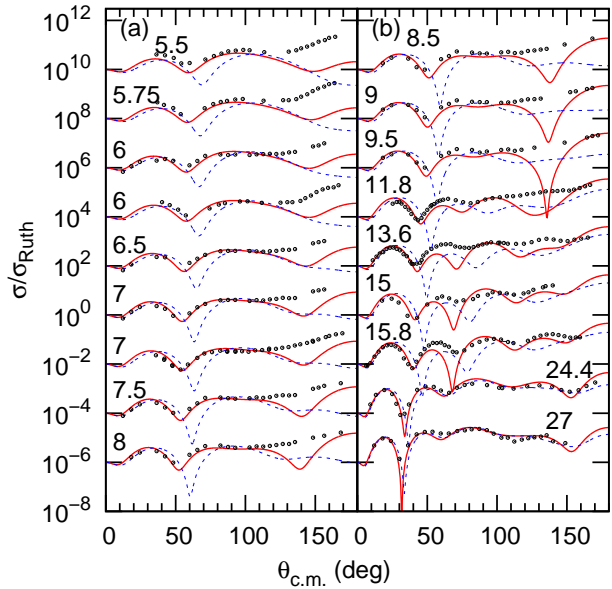}
 \caption{(Color online) Comparisons between the experimental data and optical model calculations with DA1p (solid curves) and Daehnick \textit{et al.} (dashed curves) for deuteron impinging on \nuc{9}{Be}. The deuteron incident energies are indicated along with the curves in MeV. The cross sections are offset by factors of $10^2$.}
 \label{fig-01}
\end{figure}

\begin{figure*}[htbp]
 \includegraphics[width=\textwidth]{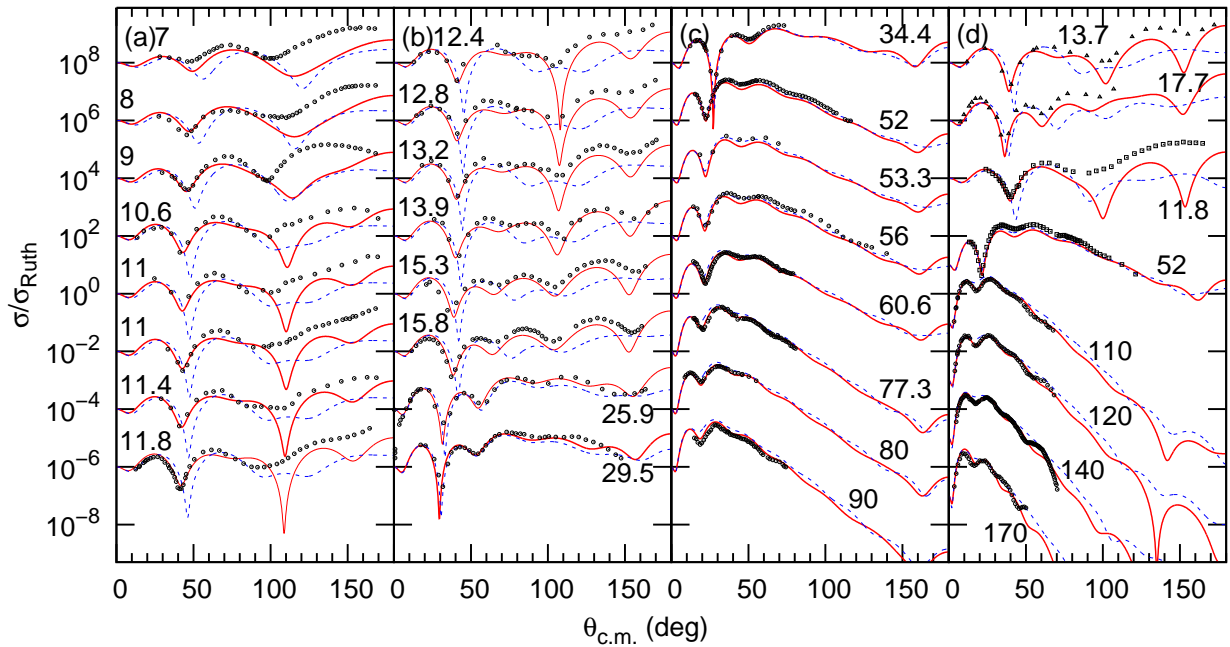}
 \caption{(Color online) The same as Fig.\ref{fig-01} but for
deuteron elastic scattering from (a,b,c) \nuc{12}{C} (circles),
(d) \nuc{13}{C} (triangles), \nuc{14}{N} (squares) and \nuc{12}{C} (circles).
 }
\label{fig-02}
\end{figure*}

\begin{figure}[htbp]
 \includegraphics[width=0.5\textwidth]{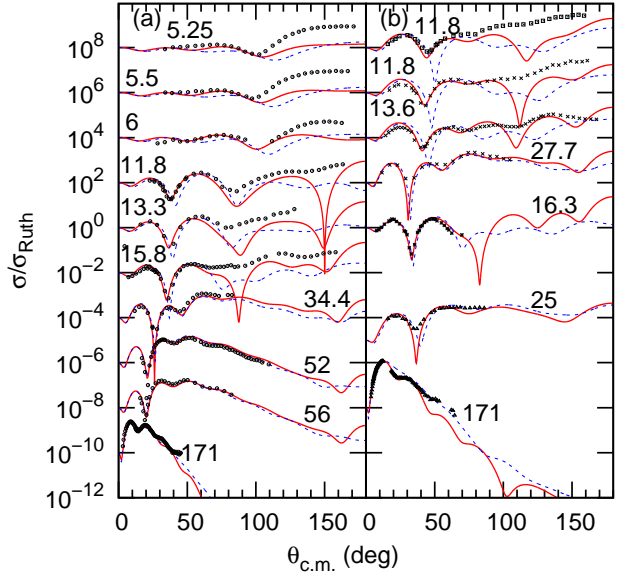}
 \caption{(Color online) The same as Fig. \ref{fig-01} but for
deuteron elastic scattering from  (a) \nuc{16}{O} (circles), (b) \nuc{10}{B} (squares), \nuc{11}{B} (X-marks), \nuc{18}{O} (asterisks) and \nuc{6}{Li} (triangles) at high energy.}
\label{fig-03}
\end{figure}

\begin{figure}[htbp]
 \includegraphics[width=0.5\textwidth]{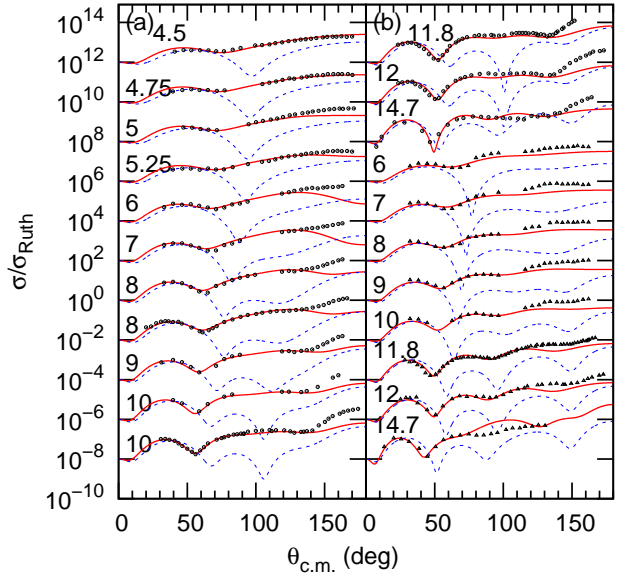}
 \caption{(Color online) The same as Fig.\ref{fig-01} but for
deuteron elastic scattering from \nuc{6}{Li} (circles) and \nuc{7}{Li} (triangles) at $E_\textrm{d}<15$ MeV.}
 \label{fig-04}
\end{figure}

\begin{figure}[htbp]
 \includegraphics[width=0.4\textwidth]{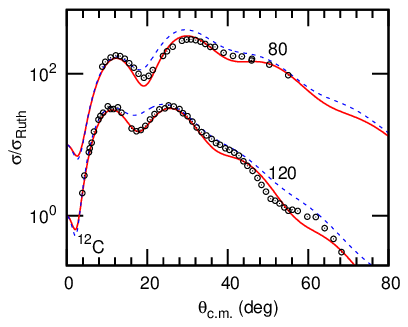}
 \caption{(Color online) Comparisons with optical model calculations and experimental data for deuteron elastic scattering from \nuc{12}{C} at 80 and 120 MeV. Separately plotted from Fig.\ref{fig-02} to show in details of comparisons between results with DA1p and the Daehnick \textit{et al.}.}
 \label{fig-05}
\end{figure}

Comparisons with experimental data and optical model calculations of the DA1p parameters are given in Figs.\ref{fig-01}-\ref{fig-04} together with the predictions using the systematics of Daehnick \textit{et al.} \cite{Daehnick}. Clearly, DA1p improves the reproduction to the experimental data with respect to that of the latter, especially at low incident energies and at forward angels. At higher incident energies above around 30 MeV, both systematic potentials give satisfactory reproduction to the experimental data. A close comparison between the optical model calculations using these two systematic potentials at higher incident energies are given in Fig.\ref{fig-05}. One sees that DA1p is better than the systematics of Daehnick \textit{et al.} in reproducing more details of the angular distributions of the experimental data. In addition, the $\chi^2$ values associated with calculations using parameters of DA1p, of Daehnick \textit{et al.}, and of Haixia An \textit{et al.} \cite{AnHX-PRC-2006} are also given in Table.\ref{tab-1p-data}. All $\chi^2$ values are calculated assuming the same experimental uncertainties. Note that DA1p does not include spin-orbit potentials while the other two systematics do. In this sense, DA1p implicitly includes the effects of the spin-orbit potential.

It is interesting to observe that the depth of the real part of DA1p, which has $V_r=98.9$ MeV,
is larger than those in the systematics established for heavy-targets, for example, the values of $V_r$
are 86, 91.85, and 82.18 MeV in the work of Daehnick \textit{et al.} \cite{Daehnick}, Haixia An \textit{et al.} \cite{AnHX-PRC-2006} and
Yinlu Han \textit{et al.} \cite{HanYL-PRC-2006}, respectively. The same differences between systematic potentials in $1p$-shell nuclei and heavy-target nuclei
are also found in systematic potentials of proton, \nuc{3}{H} and \nuc{3}{He} \cite{Watson-PR-1967,HT1p}. Also, the radius parameter of the imaginary potential, $r_\textrm{w}^{(0)}$, as shown in Table.\ref{tab-1p-parameters}, shows stronger dependence on the target masses than the systematics established in the heavy-target region. This may be related to the fact that the $1p$-shell nuclei distinguish with each other more strongly than those among the heavy targets in their structures.

\section{Application of DA1p to radiative nuclei and total reaction cross sections}\label{sec-discussions}

\begin{figure*}[htbp]
 \includegraphics[width=0.8\textwidth]{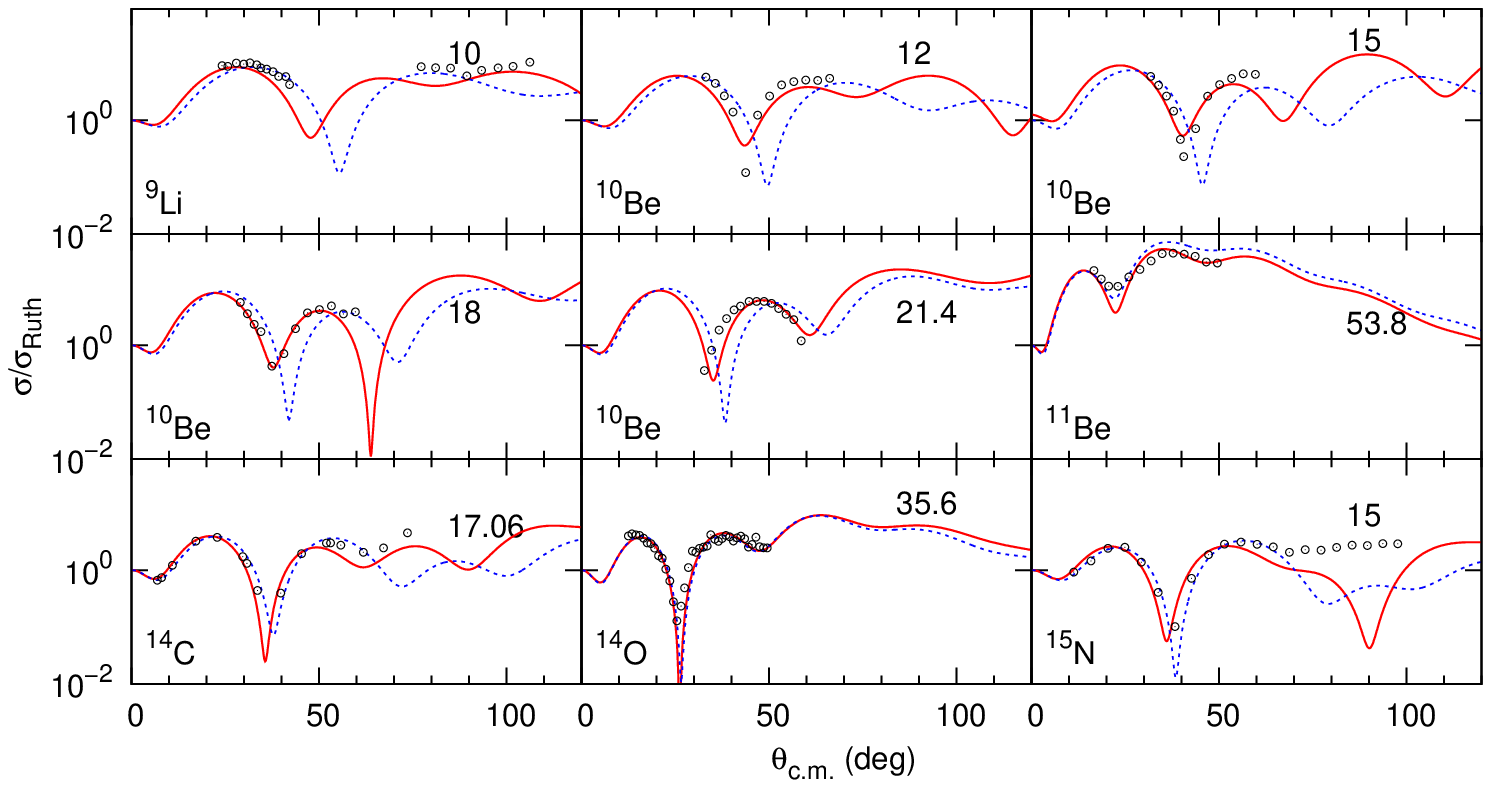}
 \caption{(Color online) The same as Fig.\ref{fig-01} but for experimental data that are not included in our systematic analysis, mostly for radiative nuclei 
 \cite{Falou-PLB-2013,Schmitt-PRC-2013,PKU-newdata,Mukhamedzhanov-PRC-2011,Flavigny-PRL-2013,GuoBing-private}.}
 \label{fig-06}
\end{figure*}

Comparisons between optical model calculations using DA1p and the experimental data which are not included in our database, 
mostly with radiative nuclei, are given in Fig.\ref{fig-06}. Again, one sees that DA1p improves the reproduction to the experimental 
data with respect to the systematic potential of Daehnick \textit{et al.}. A detailed comparison in $\chi^2$-values are given in Table.\ref{data-2}. 
This suggests that DA1p can give more reliable predictions to the elastic scattering cross sections of deuteron with nuclei that are away from the $\beta$-stability line.

\begin{table}[htbp]
\caption{The same as Table.\ref{tab-1p-data}, but for the experimental data shown in Fig.\ref{fig-06}.}
\begin{center}
\begin{tabular}{cccccccccccccc}
\hline\hline
target & $E_\textrm{d}$ & $\chi^2_{\textrm{DA1p}}$ & $\chi^2_{\textrm{Dae}}$ & $\chi^2_{\textrm{An}}$ & Ref &\\\hline
\nuc{9}{Li} & 10 & 5.11 & 4.37 &  4.21 & \cite{Falou-PLB-2013}\\
\nuc{10}{Be} & 12 & 20.61 & 262.11 & 163.01 & \cite{Schmitt-PRC-2013}\\
& 15 & 9.22 & 94.86 & 65.02 &\cite{Schmitt-PRC-2013}\\
& 18 & 3.55 & 31.73 & 25.65 &\cite{Schmitt-PRC-2013}\\
& 21.4 & 10.40 & 146.73 & 126.63 &\cite{Schmitt-PRC-2013}\\
\nuc{11}{Be} & 53.8 & 3.80 & 10.47 & 7.19 &  \cite{PKU-newdata} \\
\nuc{14}{C} &17.06 & 4.10 & 6.17 &  4.31 & \cite{Mukhamedzhanov-PRC-2011}\\
\nuc{14}{O} & 35.6 & 4.06 & 4.67 &  6.06 & \cite{Flavigny-PRL-2013}\\
\nuc{15}{N} & 15 & 14.37 & 12.87 & 13.00 & \cite{GuoBing-private}\\
\hline
\end{tabular}
\end{center}
\label{data-2}
\end{table}

Total reaction cross sections are not used to constrain the parameters of DA1p. Comparisons between optical model calculations and experimental data of 
total reaction cross sections are made for \nuc{9}{Be}, \nuc{12}{C} and \nuc{16}{O} targets for deuteron incident energies of 37.9, 65.5 
and 97.4 MeV \cite{Auce-PRC-1996}. Systematic potentials of DA1p, Haixia An \textit{et al.} and Daehnick \textit{et al.} are used here. 
These results seem to suggest that the systematics of Haixia An \textit{et al.} and Daehnick \textit{et al.} give better accounts of the total reaction cross sections
of deuterons with light targets. However, we found that the discrepencies between results with DA1p and the experimental data might be reconciled when 
the breakup of deuteron is taken into account. We will discuss this problem in details in a following paper.

\begin{figure}[htbp]
 \includegraphics[width=0.5\textwidth]{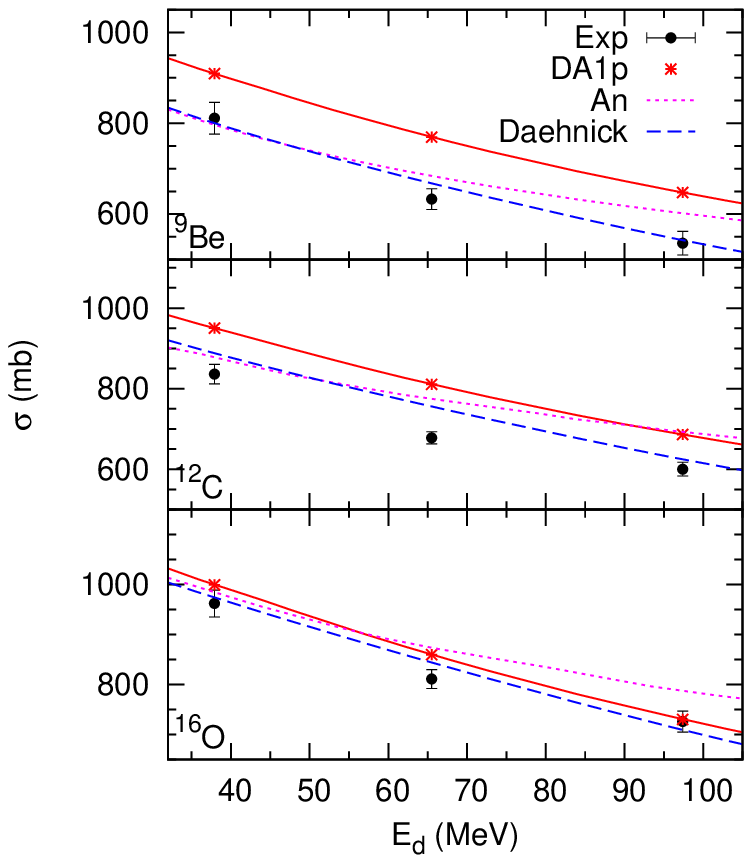}
 \caption{(Color online) Optical model calculations of the total reaction cross sections with systematics of DA1p, Daehnick \textit{et al.} and Haixia An \textit{et al.} 
 for targets \nuc{9}{Be} (upper panel), \nuc{12}{C} (middle panel) and \nuc{16}{O} (bottom panel) and their comparisons with the experimental data \cite{Auce-PRC-1996}.}
 \label{fig-07}
\end{figure}

\section{Conclusions}\label{conclusions}

In conclusion, we present in this paper a systematic phenomenological optical model potential, DA1p, of deuteron with the $1p$-shell nuclei (except for \nuc{6}{Li} and \nuc{7}{Li} ) for incident energies from  around 5 to 170 MeV. Two sets of parameters are given for \nuc{6}{Li} and \nuc{7}{Li} targets for incident energies between around 5 and 15 MeV.
Differences in the potential parameters are found between DA1p and the systematic potentials established for heavy-target region. DA1p is found to give satisfactory reproduction to the angular distributions of deuteron elastic scattering from both stable and radiative $1p$-shell nuclei.
The experimental total reaction cross sections for \nuc{9}{Be}, \nuc{12}{C} and \nuc{16}{O} targets
are found to be overpredicted by theoretical calculations with systematic deuteron potentials, which was found to be due to the breakup cross sections of deuterons and will be further studied in a following paper.

\section*{Acknowledgements}

This work is supported by the 973 Program of China(Grant No.2013CB834402), and the National Natural Science Foundation of China(Grants No.11275001, No.11275018, U1432247  and No.11535004).



\begin{thebibliography}{999}
\bibitem{LiuXD-PRC-2004}  X. D. Liu, M. A. Famiano, W. G. Lynch, M. B. Tsang, and J. A. Tostevin, Phys. Rev. C \textbf{69}, 064313 (2004).
\bibitem{Tsang-PRL-2005} M. B. Tsang, J. Lee, and W. G. Lynch, Phys. Rev. Lett. \textbf{95}, 222501 (2005).
\bibitem{Tsang-PRL-2009} M. B. Tsang, J. Lee, S. C. Su, $et\ al.$, Phys. Rev. Lett. \textbf{102}, 062501 (2009).
\bibitem{BGPN} F. D. Becchetti, Jr. and G. W. Greenless, Phys. Rev. \textbf{182}, 1190 (1969).
\bibitem{CH89} R. L. Varner, W. J. Thompson, T. L. McAbee, E. J. Ludwig and T. B. Clegg, Phys. Rep. \textbf{201}, 57 (1991).
\bibitem{KD02} A. J. Koning and J. P. Delaroche, Nucl. Phys. A \textbf{713}, 231 (2003).
\bibitem{XuRR-PRC-2012} T. Furumoto, W. Horiuchi, M. Takashina, Y. Yamamoto, and Y. Sakuragi,  Phys. Rev. C \textbf{85} 069901 (2012).
\bibitem{Perey-PRev-1963} C. M. Perey and F. G. Perey, Phys. Rev. \textbf{132}, 755 (1963).
\bibitem{Lohr-NPA-1974} J. M. Lohr and W. Haeberli, Nucl. Phys. A \textbf{232}, 381 (1984).
\bibitem{Daehnick} W. W. Daehnick, J. D. Childs, and Z. Vrcelj, Phys. Rev. C \textbf{21}, 2253 (1980).\bibitem{AnHX-PRC-2006} Haixia An and Chonghai Cai, Phys. Rev. C \textbf{73} 054605 (2006).
\bibitem{HanYL-PRC-2006} Yinlu Han, Yuyang Shi, and Qingbiao Shen, Phys. Rev. C \textbf{74}, 044615  (2006).

\bibitem{BG3He} F. D. Becchetti, Jr., and G. W. Greenlees in \textit{Polarization Phenomena in Nuclear Reactions} (H. H. Barschall and W. Haeberli, eds.) p. 682, The University of Wisconsin Press, Madison, WI. (1971).
\bibitem{GuoHR-PRC-2009} Hairui Guo, Yue Zhang, Yinlu Han, and Qingbiao Shen, Phys. Rev. C \textbf{79}, 064601 (2009).
\bibitem{GDP08} D. Y. Pang, P. Roussel-Chomaz, H. Savajols, R. L. Varner, and R. Wolski, Phys. Rev. C \textbf{79}, 024615 (2009).
\bibitem{LiXiaohua} Xiaohua Li, Chuntian Liang, Chonghai Cai, Nucl. Phys. A \textbf{789}, 103 (2007).
\bibitem{LiangChuntian} Chun-Tian Liang, Xiao-Hua Li and Chong-Hai Cai, J. Phys. G: Nucl. Part. Phys. \textbf{36}, 085104 (2009).
\bibitem{Nolte-PRC-1987}  M. Nolte, H. Machner, and J. Bojowald, Phys. Rev. C \textbf{36}, 1312 (1987).
\bibitem{Avrigeanu-ADDNT-2009} M. Avrigeanu, A. C. Obreja, F. L. Roman, V. Avrigeanu, and W. von Oertzen, At. Data Nucl. Data Tables \textbf{95}, 501 (2009).
\bibitem{Pang4He} D. Y. Pang, Y. L. Ye and F. R. Xu, Phys. Rev. C \textbf{83}, 064619 (2011).
\bibitem{Pang-JPG-2012} D. Y. Pang, Y. L. Ye, and F. R. Xu, J. Phys. G: Nucl. Part. Phys. \textbf{39}, 095101 (2012).
\bibitem{Chamon-PRC-2002} L. C. Chamon, B. V. Carlson, L. R. Gasques, $et\ al.$, Phys. Rev. C \textbf{66}, 014610 (2002).
\bibitem{Furumoto-PRC-2012} T. Furumoto, W. Horiuchi, M. Takashina, Y. Yamamoto, and Y. Sakuragi, Phys. Rev. C \textbf{85} 044607 (2012).
\bibitem{XuYP-PRC-2013} Y. P. Xu and D. Y. Pang, Phys. Rev. C \textbf{87}, 044605 (2013).
\bibitem{Johnson-Soper} R. C. Johnson and P. J. R. Soper, Phys. Rev. C \textbf{1}, 976 (1970).
\bibitem{Akram-PRC-2012}  A. M. Mukhamedzhanov, V. Eremenko, and A. I. Sattarov, Phys. Rev. C \textbf{86}, 034001 (2012).
\bibitem{Pang-PRC-2013}  D. Y. Pang, N. K. Timofeyuk, R. C. Johnson, and J. A. Tostevin, Phys. Rev. C \textbf{87}, 064613 (2013).
\bibitem{Vernotte-NPA-1994} J. Vernotte, G. Berrier-Ronsin, J. Kalifa, R. Tamisier, and B. H. Wildenthal, Nucl. Phys. A \textbf{571}, 1 (1994).
\bibitem{Jenny-CPL-2014} J. Lee, Pang Dan-Yang, Han Yin-Lu, and M. B. Tsang, Chin. Phys. Lett. \textbf{31}, 092103 (2014).

\bibitem{ChenJie-PRC-2016} J. Chen, J. L.  Lou, Y. L. Ye, et al., Phys. Rev. C \textbf{93}, 034623 (2016).
\bibitem{Watson-PR-1967}  B. A. Watson, P. P. Singh and R. E. Segel, Phys. Rev. \textbf{182}, 4 (1969).
\bibitem{HT1p} D. Y. Pang, W. M. Dean A. M. Mukhamedzhanov,  Phys. Rev. C \textbf{91}, 024611 (2015).
\bibitem{Faisal-CPL-2010} Faisal Jamil-Qureshi, Lou Jian-Ling, Ye Yan-Lin, $et\ al.$, Chin. Phys. Lett. \textbf{27}, 092501 (2010).
\bibitem{PangandYe-PRC-2011} D. Y. Pang, Y. L. Ye and F. R. Xu, Phys. Rev. C \textbf{83}, 064619 (2011).
\bibitem{FRESCO}  I. J. Thompson, Comput. Phys. Rep. \textbf{7}, 167 (1988).
\bibitem{exfor} N. Otuka, E. Dupont, V. Semkova $et\ al.$, Nucl. Data Sheets \textbf{120}, 272 (2014).
\bibitem{Efron-bootstrap} B. Efron, Biometrika 68 (1981) 589; ibid., SIAM Rev. 21(1979) 460, and P. Diaconis and B. Efron, Sci. Am. 248, 116 (1983).


\bibitem{POWELL-NPA-1970} D. L. Powell, G. M. Crawley, B. V. N. Rao, $et\ al.$, Nucl. Phys. A \textbf{147}, 65 (1970).
\bibitem{SZCZUREK-ZPA-1989} A. Szczurek, K. Bodek, J. Krug, $et\ al.$, Z. Phys. A \textbf{333}, 271 (1989). 
\bibitem{Dantzig-NP-1963} R. Van Dantzig, L. A. Ch. Koerts, Nucl. Phys. \textbf{48}, 177 (1963).

\bibitem{Generalov-IZV-2000} L. N. Generalov, S. N. Abramovich, A. G. Zvenigorodskij,  
Jour. Izv. Rossiiskoi Akademii Nauk, Ser.Fiz. \textbf{64}, 440 (2000).
\bibitem{Newman-NPA-1967} E. Newman, L. C. Becker, B. M. Preedom, Nucl. Phys.  A \textbf{100}, 225 (1967).
\bibitem{Hinterberger-NPA-1968} F. Hinterberger, G. Mairle, U. Schmidt-Rohr, $et\ al.$, Nucl. Phys.  A \textbf{111}, 265 (1968).
\bibitem{Abramovich-IZV-1976} S. N. Abramovich, B. Ja. Guzhovskij, B. M. Dzuba, $et\ al.$, Jour. Izv. Rossiiskoi Akademii Nauk, Ser.Fiz.
\textbf{40}, 842 (1976).
\bibitem{Ishida-PLB-1993} S. Ishida, H. Sakai, H. Okamura, Phys. Lett. B \textbf{314}, 279 (1993).
\bibitem{Perrin-NPA-1977} G. Perrin, Nguyen Van Sen, J. Arvieux, $et\ al.$, Nucl. Phys. A \textbf{282}, 221 (1977).
\bibitem{Matsuoka-NPA-1986} N. Matsuoka, H. Sakai, T. Saito, $et\ al.$, Nucl. Phys. A \textbf{455}, 413 (1986).
\bibitem{Fitz-NPA-1967} W. Fitz, R. Jahr, R. Santo, Nucl. Phys. A \textbf{101}, 449 (1967).
\bibitem{BINGHAM-NPA-1971} H. G. Bingham, A. R. Zander, K. W. Kemper, $et\ al.$, Nucl. Phys. A \textbf{173}, 265 (1971).
\bibitem{Vereshchagin-SPJ-1970} A. N. Vereshchagin, I. N. Korostova, I. P. Chernov,  
Jour. Izv. Rossiiskoi Akademii Nauk, Ser.Fiz.,  \textbf{32}, 623 (1968).
\bibitem{DARGEN-NPA-1976} S. E. Darden, G. Murillo, S. Sen, Nucl. Phys. A \textbf{266}, 29 (1976).
\bibitem{Duhamel-NPA-1971} G. Duhamel, L. Marcus, H. Langevin-Joliot, $et\ al.$, Nucl. Phys. A \textbf{174} 485 (1971).
\bibitem{COWLEY-NP-1966} A. A. Cowley, G. Heymann, R. L. Keizer, $et\ al.$, Nucl. Phys. \textbf{86}, 363 (1966).
\bibitem{SUMMERSGILL-PRev-1958} R. G. Summers-Gill, Phys. Rev. \textbf{109}, 1591 (1958).
\bibitem{Betker-PRC-1993} A. C. Betker, C. A. Gagliardi, D. R. Semon, $et\ al.$, Phys. Rev. C \textbf{48}, 2085 (1993).
\bibitem{Ludecke-NPA-1968} H. L\"{u}decke, Tan Wan-Tjin, H. Werner and J. Zimmerer, Nucl. Phys. A \textbf{109}, 676 (1968)
\bibitem{Slobodrian-NP-1962} R. J. Slobodrian, Nucl. Phys. \textbf{32}, 684 (1962).
\bibitem{Okamura-PRC-1998} H. Okamura, S. Ishida, N. Sakamoto, $et\ al.$, Phys. Rev. C \textbf{58}, 2180(1998).
\bibitem{MATSUKI-JPJ-1969} S. Matsuki, S. Yamashita, K. Fukunaga, $et\ al.$, J. Phys. Soc. Japan \textbf{26}, 1344 (1969).
\bibitem{Aspelund-NPA-1975} O. Aspelund, G. Hrehuss, A. Kiss, $et\ al.$, Nucl. Phys. A \textbf{253}, 263 (1975).
\bibitem{Baeumer-PRC-2001} C. B\"{a}umer, R. Bassini, A. M. van den Berg, $et\ al.$, Phys. Rev. C \textbf{63}, 037601 (2001).
\bibitem{Burtebyaev-PAN-2010} N. Burtebayev, S. V. Artemov, B. A. Duisebayev, Phys. At. Nucl. \textbf{73}, 746 (2010).
\bibitem{Guratzsch-NPA-1970} H. Guratzsch, J. Slotta, G. Stiller, Nucl. Phys. A \textbf{140}, 129 (1970).
\bibitem{Korff-PRC-2001} A. Korff, P. Haefner, C. B\"{a}umer, $et\ al.$, Phys. Rev. C \textbf{70}, 067601 (2004).
\bibitem{Peterson-NPA-1984} R. J. Peterson, H. C. Bhang, J. J. Hamill, $et\ al.$, Nucl. Phys. A \textbf{425}, 469 (1984).
\bibitem{Ohlsen-NP-1963} G. G. Ohlsen, R. E. Shamu, Nucl. Phys. \textbf{45}, 523 (1963).
\bibitem{Davison-CJP-1970} N. E. Davison, W. K. Dawson, G. Roy, $et\ al.$, Canadian Journal of Physics, \textbf{48}, 2235 (1970).
\bibitem{BALDEWEG-NP-1966} F. Baldeweg, V. Bredel, H. Guratzsch, $et\ al.$, Nucl. Phys. \textbf{84}, 305 (1966).
\bibitem{Corrigan-NPA-1972} K. W. Corrigan, R. M. Prior, S. E. Darden, $et\ al.$, Nucl. Phys. A \textbf{188}, 164 (1972).
\bibitem{Hatanaka-NPA-1980} K. Hatanaka, K. Imai, S. Kobayashi, $et\ al.$, Nucl. Phys. A \textbf{340}, 93 (1980).
\bibitem{Galanina-PAN-2007} L. I. Galanina, N. S. Zelenskaya, V. M. Lebedev, $et\ al.$, Physics of Atomic Nuclei, \textbf{70}, 273 (2007).
\bibitem{Burjan-JP} V. Burjan, Z. Hons, V. Kroha, $et\ al.$, Journal of Physcis: Conference Series, \textbf{420}, 012142 (2013).
\bibitem{Auce-PRC-1996} A. Auce, R. F. Carlson, A. J. Cox, $et\ al.$, Phys. Rev. C \textbf{53}, 2919(1996).

\bibitem{Falou-PLB-2013} H. Al Falou, R. Kanungo, C. Andreoiu, $et\ al.$,  Phys. Lett. B \textbf{721}, 224(2013).
\bibitem{Schmitt-PRC-2013} K. T. Schmitt, K. L. Jones, S. Ahn, $et\ at.$, Phys. Rev. C \textbf{88}, 064612(2013).
\bibitem{PKU-newdata} Result from the same experiment as in Ref.\cite{ChenJie-PRC-2016}, to be published.
\bibitem{Mukhamedzhanov-PRC-2011} A. M. Mukhamedzhanov, V. Burjan, M. Gulino, et. al., Phys. Rev. C \textbf{84}, 024616(2011).
\bibitem{Flavigny-PRL-2013} F. Flavigny, A. Gillibert, L. Nalpas, et. al., Phys. Rev. Lett. \textbf{110}, 122503(2013).
\bibitem{GuoBing-private} B. Guo, private communications.

\end{thebibliography}
\end{document}